\documentclass[cup7b]{cupbook}
\usepackage{graphicx,epstopdf}
\usepackage{amssymb,amsmath,epsfig}
\newcommand{\beq}{\begin{equation}}
\newcommand{\eeq}{\end{equation}}
\newcommand{\barr}{\begin{array}}
\newcommand{\earr}{\end{array}}

\newcommand{\Hi}{\mathcal{H}}
\newcommand{\Ai}{\mathcal{A}}
\newcommand{\m}{\Phi}

\def\A{{\cal A}}

\def\K{{\cal K}}

\def\P{{\mathcal P}}

\def\S{{\cal S}}

\newtheorem{definition}{Definition}

\newcommand{\G}{\Gamma}

\newcommand{\Hc}{{\cal H}}
\newcommand{\Aevol}{{\cal A}_{\mbox{\footnotesize{ evol}}}}

\begin{document}
\author[F. Markopoulou]{Fotini Markopoulou\\ Perimeter Institute for Theoretical Physics}

\chapter[New directions in Background Independent Quantum Gravity]{New directions\\ in Background Independent\\ Quantum Gravity}

\date{March 19, 2007}

\begin{abstract}
We discuss the meaning of background independence in quantum theories of gravity where geometry and gravity are emergent and illustrate the possibilities using the framework of quantum causal histories.
\end{abstract}

\section{Introduction}

The different approaches to quantum gravity can be classified according to the role that spacetime plays in them.  In particular, we can ask two questions of each approach:  1) Is spacetime geometry and general relativity fundamental or emergent? 2) Is spacetime geometry, if present, dynamical or fixed?  

Reviewing the different approaches we find that they split into four categories.  First, there are the quantum field theory-like approaches, such as string theory and its relatives.  Here general relativity is to be an emergent description, however, the spacetime that appears in the initial formulation of the theory is fixed and not dynamical.  Next are the so-called background independent approaches to quantum gravity, such as loop quantum gravity, spin foams, causal sets and causal dynamical triangulations.  Geometry and gravity here is fundamental, except quantum instead of classical.  These approaches implement background independence by some form of superposition of spacetimes, hence the geometry is not fixed.  Third, there are condensed matter approaches (see Volovik, 2006). While it is clear that relativity is to be emergent, there is confusion on question 2 above.  These are condensed matter systems, so it seems clear that there is a fixed spacetime in which the lattice lives, however, it can be  argued that it is an auxilliary construction, an issue we shall not resolve here.  

Our main focus in this chapter is a new, fourth, category that is currently under development and constitutes a promising and previously unexplored direction in background independent quantum gravity.  This is {\em pre-geometric} background independent approaches to quantum gravity.  These start with an underlying microscopic theory of quantum systems in which no reference to a spatiotemporal geometry is to be found.  Both geometry and hence  gravity are  emergent.  The geometry is defined intrinsically using  subsystems and their interactions.   The geometry is  subject to the dynamics and hence itself dynamical.  This has been claimed to be the case, in different systems, by Dreyer 2004 and this volume; Lloyd, 2005, Kribs and Markopoulou, 2005 and Konopka, Markopoulou and Smolin, 2006.

As can be seen from the above, this new direction is in fact orthogonal to all previous approaches and so it comes with its own set of promises and challenges.  We shall discuss these but we also wish to outline the choices involved in the answers to our two questions above.  It is normally difficult to have an overview of the choices involved in picking different directions in quantum gravity because the mathematical realizations are intricate and all different.  Luckily, for the present purposes, we find that we can base the discussion on the formalism of {\em Quantum Causal Histories} (QCH), a locally finite directed graph of finite-dimensional quantum systems\footnote{The finiteness is a simple implementation of the expectation that there really are only a finite number of degrees of freedom in a finite volume, arguments for which are well-known and we have reviewed them elsewhere (Markopoulou, 2002).}.  

A QCH, depending on the physical interpretation of its constituents, can model a discrete analogue of quantum field theory, a traditional, quantum geometry based background independent system, or the new, pre-geometric background independent theories.   This will allow us to keep a overview of the forks on the road to quantum gravity.  It will also be ideal for analyzing the newest kind of background independent systems and obtaining some first results on their effective properties.  In particular, we shall see how one can extract conserved quantities in pre-geometric systems using a straightforward map between a QCH and a quantum information processing system.  

The outline of this chapter is as follows.  In section \ref{QCH} we give the definition of a Quantum Causal History, together with a simple example, locally evolving networks in subsection \ref{ex}.  At this point we have not restricted ourselves to any particular physical interpretation of the QCH and the options are listed in \ref{Gamma}.  In section \ref{BI} we give the necessary definitions of background independence.  The following three sections contain three distinct physical interpretations of a QCH:  As a discrete analogue of quantum field theory (a background-dependent theory) in section \ref{QFT}, a quantum geometry theory in section \ref{QG} with a discussion of advantages and challenges (\ref{ACQG}) and finally the new type of background independent systems in section \ref{pregeo}.  Their advantages and challenges are discussed in \ref{ACpregeo}.  In section \ref{NS},  we map a QCH to a quantum information processing system and use this to derive conserved quantities with no reference to a background spacetime, complete with a simple example of such conserved quantities.  We conclude with a brief discussion of these new directions in section \ref{conc}.

%%%%%%%%%%%%%%%%%%%%%%%%%%%%%%%%%%%%%%%%%%%%%

\section{Quantum Causal Histories}\label{QCH}

A quantum causal history is a locally finite directed graph of finite-dimensional quantum systems.  We start by giving the properties of the directed graph and the assignment of quantum systems to its vertices and appropriate operators to its edges. 
The addition of 3 axioms
ensures that the properties of a given graph are reflected in the flow of physical information in the corresponding quantum operators
and completes the definition of a Quantum Causal History
\footnote{The abstract form of a Quantum Causal History based on a directed graph that we follow here was given by Kribs, 2005, based on the original definition in Markopoulou, 2000 and Hawkins, Markopoulou and Sahlmann, 2003.}.

Let $\Gamma$ be a directed graph with vertices $x\in V(\Gamma)$
and directed edges $e\in E(\Gamma)$. The {\it source} $s(e)$  and
{\it range} $r(e)$ of an edge $e$ are, respectively, the initial
and final vertex of $e$. A (finite) path $w=e_k \cdots e_1$ in
$\Gamma$ is a sequence of edges of $\Gamma$ such that $r(e_i) =
s(e_{i+1})$ for $1\leq i < k$. If $s(w) = r(w)$ then we say $w$ is
a {\it cycle}. We require that $\Gamma$ has no
cycles.

If there exists a
path $w$ such that   $s(w)=x$ and $r(w)=y$
 let us write $x\leq y$ for
the associated partial ordering.  We call such vertices {\em related}.  Otherwise, they are {\em unrelated}.  We use $x\sim z$ to denote that $x$ and $z$ are unrelated.  Given any $x\leq y$, we require that there are finitely many $z\in V(\Gamma)$
such that $x\leq z \leq y$.  This is the condition of {\em local finiteness}.

\begin{definition}
Parallel set, complete source, complete range, complete pair.

A {\em parallel set} $\xi\subseteq E(\Gamma)$ is defined by the
property that $x\sim y$ whenever $x,y\in\xi$. A parallel set $\xi$ is a {\em complete source} of $x$ if all paths $w$ with $r(w)\equiv x$ have $s(w)\in\xi$.  Conversely, a parallel set $\zeta$ is a {\em complete range} of $x$ if all paths $w$ with source $s(w)\equiv x$ have range in $\zeta$, $r(w)\in\zeta$.  Two parallel sets $\xi$ and $\zeta$ are a {\em complete pair} if all paths $w$ that start in $\xi$  $s(w)\in\xi$ end up in $\zeta$, $r(w)\in\zeta$ and the reverse. 
\end{definition}
 For example, in the directed graph 
\[
	\begin{array}{c}\mbox{\includegraphics[height=3cm]{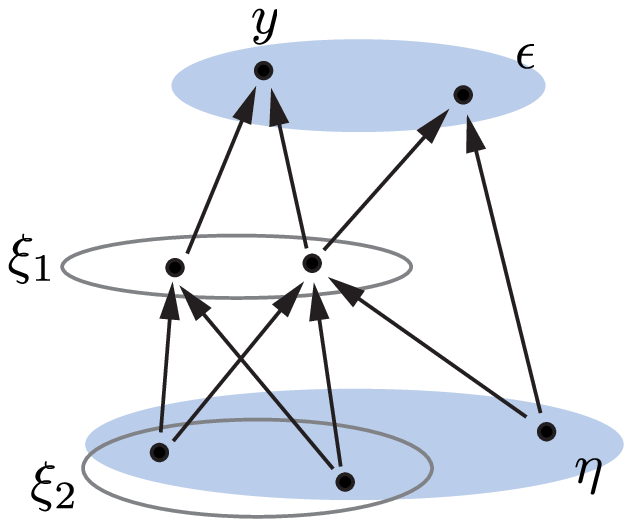}}\end{array}
\]
$\xi_1$ is a complete source  for $y$ while the parallel sets $\xi_2$ and  $\eta$ are  not.  The sets $\eta$ and $\epsilon$ are a complete pair.

We now wish to associate quantum systems to the graph.  
The construction of a  quantum causal history starts with a  directed graph $\Gamma$
and assigns to every
vertex $x\in V(\Gamma)$ a finite-dimensional Hilbert space
$\Hi(x)$  and/or a matrix algebra $\Ai(\Hi(x))$ (or $\Ai(x)$ for short) of operators acting on $\Hi(x)$.  It is best to regard the algebras as the primary objects, but we will not make this distinction here.  

If two vertices, $x$ and $z$, are unrelated, their joint state space is 
\beq
\Hi\left(x\cup z\right)=\Hi(x)\otimes\Hi(y).
\eeq
If vertices $x$ and $y$ are related, let us for simplicity say by a single edge $e$, 
we shall think of $e$ as a {\em change} of the quantum systems of the source of $e$ into a new set of quantum systems (the range of $e$).  It is then natural to assign to each $e\in E(\Gamma)$ a {\em completely positive map} $\Phi_e$:
\beq
\Phi_e: {\cal A}(s(e))\longrightarrow{\cal A}(r(e)),\label{eq:qchphi}
\eeq
where ${\cal A}(x)$ is the full matrix algebra on $\Hi(x)$.  Completely positive maps are commonly used to describe evolution of open quantum systems and generally arise as follows (see, for example, Nielsen and Chuang, 2000).

Let $\Hi_S$ be the state space of a quantum system in contact with
an environment $\Hi_E$ (here $\Hc_S$ is the subgraph space and $\Hc_E$ the space of the rest of the graph). 
The standard characterization of evolution
in open quantum systems starts with an initial state in the system
space that, together with the state of the environment, undergoes
a unitary evolution determined by a Hamiltonian on the composite
Hilbert space $\Hi = \Hi_S \otimes \Hi_E$, and this is followed by
tracing out the environment to obtain the final state of the
system. 

The associated evolution map
$\Phi:{\cal A}(\Hi_S)\rightarrow{\cal A}(\Hi_S)$ between the corresponding matrix algebras of operators on the respective Hilbert spaces  is necessarily completely
positive (see below) and trace preserving. More generally, the map
can have different domain and range Hilbert spaces. Hence the
operational definition of quantum
evolution $\Phi$ from a Hilbert space $\Hi_1$ to
$\Hi_2$ is:
\begin{definition} Completely positive (CP) operators.
A {\em completely positive} operator  $\Phi$ is a linear map 
$\Phi: {\cal A}(\Hi_1)\longrightarrow{\cal A}(\Hi_2)$ such that the maps
\beq
 id_k \otimes \Phi :  M_k \otimes {\cal A}(\Hi_1) \rightarrow
 M_k \otimes {\cal A}(\Hi_2)
\eeq
are positive for all $k\geq 1$. 
\end{definition}
Here we have written $M_k$ for
the algebra ${\cal A}(\mathbb{C}^k)$.

Consider vertices $x,y,z$ and $w$ in $\G$.  There are several possible connecting paths, such as
\[
	\begin{array}{c}\mbox{\includegraphics[height=2cm]{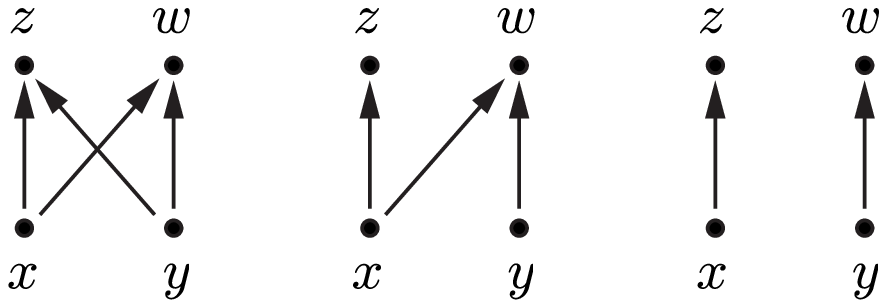}}\end{array}.
\]
We need the quantum evolution from $\A(x\cup y)$ to $\A(z\cup w)$ to reflect the underlying graph configuration (the quantum operators should distinguish between the above diagrams).  The following definition ensures this.

\begin{definition}
    A {\em quantum causal history} consists of a simple matrix algebra $\A(x)$ for every vertex $x\in V(\G)$ and a completely positive map $\m(x,y) : \A(y)\to\A(x)$ for every pair of related vertices $x\leq y$, satisfying the following axioms.
\end{definition}

\begin{tabular}{lc}
\multicolumn{2}{l}{Axiom 1: Extension.}\\
\\
\parbox{6.5cm}{\sloppy 
Let $\xi$ be the complete source of $y$ and $x\in\xi$.  For any such $y$, there exists a homomorphism $\Phi(\xi,y):\A(y)\rightarrow\A(\xi)$ such that the reduction of $\Phi(\xi,y)$ to $\A(y)\rightarrow\A(x)$ is $\Phi(x,y)$}
&
$\begin{array}{c}\mbox{\includegraphics[height=2cm]{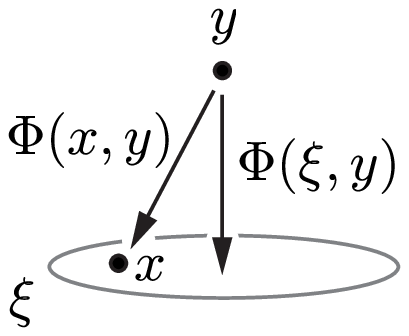}}\end{array}$\\
\parbox{6.5cm}{\sloppy 
Similarly, for the reflected diagram on the right for $\zeta$ a complete range of $y$.  The adjoint of $\Phi(y,\zeta)$ is a homomorphism while 
its reduction to $y\rightarrow z$ is $\Phi(y,z)$.
 }
&
$\begin{array}{c}\mbox{\includegraphics[height=2cm]{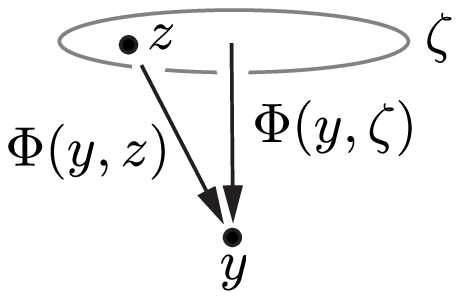}}\end{array}$
%\\
%\\
\end{tabular}
\newpage
\begin{tabular}{lc}
\multicolumn{2}{l}{Axiom 2:  Commutativity of unrelated vertices. }\\
\\
\parbox{6.5cm}{\sloppy 
If $x\sim z$ and $\xi$ is a complete source of both $y$ and $z$, then the images of $\Phi(\xi,z)$ and $\Phi(\xi,y)$ in $\A(\xi)$ commute.  
 }
&
$\begin{array}{c}\mbox{\includegraphics[height=2cm]{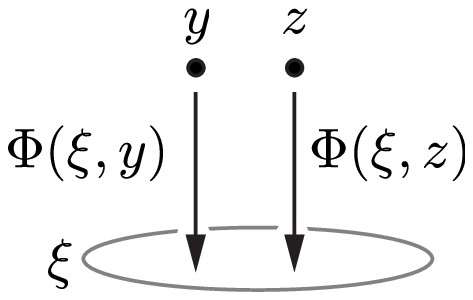}}\end{array}$
\\
\parbox{6.5cm}{\sloppy 
Similarly, on the right,  the images of $\Phi^{\dagger}(x,\zeta)$ and $\Phi^{\dagger}(y,\zeta)$ in $\A(\zeta)$ commute.  
 }
&
$\begin{array}{c}\mbox{\includegraphics[height=2cm]{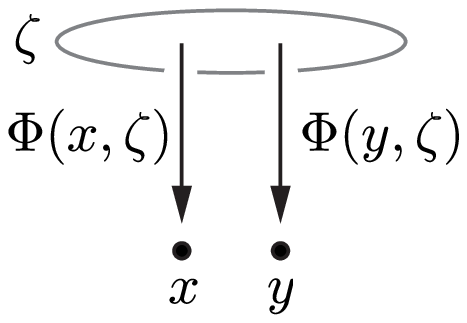}}\end{array}$
\\
\multicolumn{2}{l}{Axiom 3: Composition.}\\
\\
\parbox{6.5cm}{\sloppy 
If $\xi$ is a complete source of $z$ and a complete range of $y$, then $\Phi(y,z)=\Phi(y,\xi)\circ\Phi(\xi,z)$.\\
Similarly for the reverse direction. 
 }
&
$\begin{array}{c}\mbox{\includegraphics[height=3.6cm]{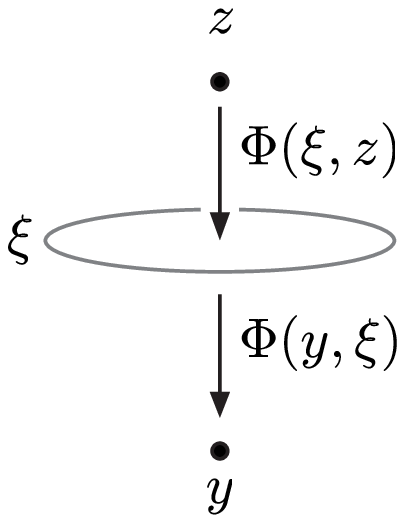}}\end{array}$
\end{tabular}\\
Note that completely positive maps between algebras go in the reverse direction to the edges of the graph.  This is as usual for maps between  states (forward) and between operators (pullbacks).

The above axioms ensure that the actual relations between the vertices of a given graph are reflected in the operators of the QCH\footnote{Very interesting recent results of Livine and Terno (2006) further analyze and constrain the allowed graph structure to take into account the quantum nature of the physical information flow represented. }.   Furthermore, as shown in Hawkins, Markopoulou and Sahlmann, if we are given the CP maps on the edges, these axioms mean that {\em unitary} operations will be found at the right places:  interpolating between {\em complete pairs}.  
When $\xi$ and $\zeta$ are a complete pair, we can regard the subgraph that interpolates between $\xi$ and $\zeta$ as the evolution of an isolated quantum system.  We would expect that in this case the composite of the individual maps between $\xi$ and $\zeta$ is unitary and indeed the above axioms ensure that this is the case.  

%%%%%%%%%%%%%%%%%%%%%%%%%%%%%%%%%%%%%%%%%%%
\subsection{Example:  Locally evolving networks of quantum systems}\label{ex}

Possibly the most common objects that appear in background independent theories are networks.  Network-based, instead of metric-based, theories are attractive implementations of the relational content of diffeomorphism invariance:  it is the connectivity of the network (relations between the constituents of the universe) that matter,  not their distances or metric attributes.
We shall use a very simple  network-based system as a concrete example of a QCH.  

We start with a network $\S$ of $n=1,\ldots,N$ nodes, each with three edges attached to it, embedded in a topological 3-dimensional space $\Sigma$ (no metric on $\Sigma$).  The network $\S$ is not to be confused with the graph $\G$, it is changes of $\S$ that will give rise to $\G$.  A map from $\S$ to a quantum system can be made by associating a finite-dimensional state space $\Hc_n$ to each minimal piece of $\S$, namely,  one node and three open edges:
\beq
	\Hc_n=\begin{array}{c}\mbox{\includegraphics{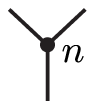}}\end{array}.
\eeq
Two such pieces  of $\S$ with no overlap are unrelated and thus the state space of the entire network $\S$ is the tensor product over all the constituents,
\beq
\Hc_\S=\bigotimes_{n\in\S}\Hc_n,
\label{eq:HG}
\eeq
and the state space of the theory is 
\beq
\Hc=\bigoplus_{\S_i}\Hc_{\S_i},
\label{eq:H}
\eeq
where the sum is over all topologically distinct embeddings of all such networks in
$\Sigma$ with the natural inner product $\langle\S_i|\S_{i'}\rangle=\delta_{\S_i\S_{i'}}$.

\begin{figure}
\begin{center}
\begin{equation}
\begin{array}{rl}
A_1&=\begin{array}{c}\includegraphics{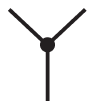}\end{array}
\longrightarrow
\begin{array}{c}\includegraphics{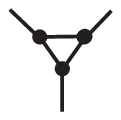}\end{array}
\\
A_2&=\begin{array}{c}\includegraphics{Hntriangle.eps}\end{array}
\longrightarrow
\begin{array}{c}\includegraphics{Hn1.eps}\end{array}
\\
A_3&=\begin{array}{c}\includegraphics{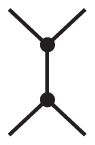}\end{array}
\longrightarrow
\begin{array}{c}\includegraphics{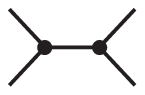}\end{array}
\end{array}
\end{equation}
\end{center}
\caption{
The three generators of evolution on the network space $\Hc$. They are called expansion,
contraction and exchange moves.}
\label{fig:moves}
\end{figure}

Local dynamics on ${\cal H}$ can be defined by excising pieces  of $\S$ and replacing them with new ones with the same boundary (Markopoulou,1997; Markopoulou and Smolin,1997).  The generators of such dynamics are given graphically in Fig.\ref{fig:moves}. 
 Given a  network $\S$, application of $A_i$ results in 
 \beq
\hat{A}_i |\S \rangle = \sum_{\alpha} |\S^\prime_{\alpha i}\rangle,
\eeq
 where $\S^\prime_{\alpha i}$ are all the networks obtained
 from $\S$ by an application of one move of type $i$  ($i=1,2,3$).  
Together with the identity ${\bf 1}$, these moves generate the {\em evolution algebra}
\beq
\Aevol=\left\{{\bf 1},A_i\right\}, \qquad i=1,2,3
\eeq
 on ${\cal H}$ . 

Finally, changing the network $\S$ by the above local moves produces a directed graph $\G$.  The vertices of $\S$ are also the vertices of $\G$.  The generator moves correspond to complete pairs and hence unitary operators, however, the operators between individual vertices are CP and the resulting system of locally evolving networks is a Quantum Causal History.  For example, in this change of $\S$ to $\S'$
\[
	\begin{array}{c}\mbox{\includegraphics[height=3cm]{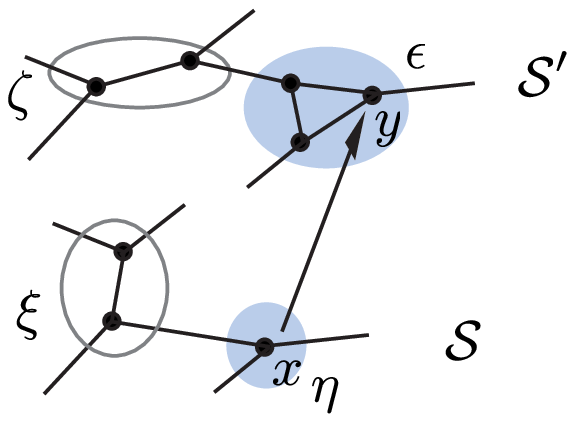}}\end{array}
\]
we have operated with $A_3$ between complete pair sets $\xi$ and $\zeta$ and with $A_1$ between complete pairs $\eta$ and $\epsilon$.  The map from $x$ to $y$ is a CP map.

%%%%%%%%%%%%%%%%%%%%%%%%%%%%%%%%%%%%%%%%%%%%%%%%%%%

\subsection{The meaning of $\G$.}\label{Gamma}

At this stage we have said nothing about the physical interpretation of $\G$ or the individual quantum systems $\A(x)$ on its vertices.  While $\G$ has the same properties as a causal set\footnote{Bombelli, Lee, Meyer and Sorkin, 1987.}, i.e., the discrete analogue of a Lorentzian spacetime, it does not have to be one.  For example, in the circuit model of quantum computation, a {\em circuit}, that is, a collection of gates and wires  also has the same properties as $\G$ and simply represents a sequence of information transfer which may or  may not be connected to spatiotemporal motions
(see Nielsen and Chuang, 2000, p.\ 129).  

We shall use this flexibility of the QCH to illustrate both the difference between a background dependent and a background independent system as well as  the distinction between background independent theories of quantum geometry and a new set of pre-geometric theories that have been recently proposed.  In what follows, we shall see that three different interpretations of $\G$ and the $\A(x)$'s give three different systems:  
1) A discrete version of algebraic quantum field theory, when $\G$ is a discretization of a Lorentzian spacetime
and $\A(x)$ is matter on it.
2) A causal spin foam, i.e., a background independent theory of quantum geometry.  Here $\G$ is a locally finite analogue of a Lorentzian spacetime and the $\A(x)$ contain further quantum geometric degrees of freedom.  Such a theory is background independent when we consider a quantum superposition of all $\G$'s.  
3) A pre-geometric background independent theory, when neither $\G$ nor the $\A(x)$'s have geometric information.  The possibility that such a system, with a single underlying graph $\G$ may be background independent has only recently been raised and explored. 

We shall discuss each of these three possibilities in detail in the rest of this chapter, starting with the necessary definitions of background independence, next.

%%%%%%%%%%%%%%%%%%%%%%%%%%%%%%%%%%%%%%%%%%%%%

\section{Background Independence}\label{BI}

Background independence (BI) is thought to be an important part of a quantum theory of gravity since it is an important part of the classical theory\footnote{
Butterfield and Isham, 2000; Stachel, 2005; Smolin, 2005.}.  Background independence in general relativity is the fact that physical quantities are invariant under spacetime diffeomorphisms.  There is no definite agreement on the form that BI takes in quantum gravity.  Stachel gives  the most concise statement of background independence:  ``In a background independent theory there is no kinematics independent of dynamics''.  

In the present article, we shall need to discuss specific aspects of background independence and to aid clarity we give the following definitions that we shall use:

\begin{definition}
Background independence I (BI-I):
A theory is background independent if its basic quantities and concepts do not presuppose the existence of a given background spacetime metric.  
\end{definition}

All well-developed background independent approaches to quantum gravity such as Loop Quantum Gravity\footnote{Ashtekar 1988;
Rovelli, 2000; Thiemann, 2007, Smolin, 2004.
}, causal sets\footnote{
Bombelli, Lee, Meyer and Sorkin, 1987.
}, spin foams\footnote{
Reisenberger, 1994; Reisenberger and Rovelli, 1997; Markopoulou and Smolin, 1997; Baez, 1998; Oriti, 2001 and this volume. 
}, causal dynamical triangulations\footnote{
Ambj{\o}rn and Loll, 1998; Ambj{\o}rn, Jurkiewicz and Loll, 2002; 2004.
}, dynamical triangulations or quantum Regge calculus\footnote{
Regge and Williams, 2000.
} implement background independence as a special case of the above by quantum analogy to the classical theory:

\begin{definition}
Background independence II (BI-II):  A background independent theory of quantum geometry is characterized
by a) quantum geometric microscopic degrees of freedom or a regularization of the microscopic geometry and b) a quantum sum-over-histories of the allowed microscopic causal histories (or equivalent histories in the Riemannian approaches).  
\end{definition}

Recently, new approaches to quantum gravity have been proposed that satisfy BI-I but not BI-II:  the computational universe (Lloyd, 2005), internal relativity (Dreyer, 2004 and this volume) and quantum graphity (Konopka, Markopoulou and Smolin, 2006).   More specifically, Dreyer advocates that

\begin{definition} Background Independence (Dreyer):  A theory is background independent if all observations are {\em internal}, i.e., made by observers inside the system.
\end{definition}
Note that this is a natural condition for a cosmological theory as has also been pointed out in Markopoulou, 2000. 

In summary, what constitutes a background independent theory is a question that is currently being revisited and  new, on occasion radical, suggestions have been offered.  These are opening up new exciting avenues in quantum gravity research and will be our focus in this chapter.  In order to discuss them in some detail, however, we shall give examples of each in the unifying context of QCH.  

%%%%%%%%%%%%%%%%%%%%%%%%%%%%%%%%%%%%%%%%%%%%%%

\section{QCH as a discrete quantum field theory}\label{QFT}

There is substantial literature in quantum gravity and high energy physics that postulates that in a finite region of the universe there should be only a finite number of degrees of freedom, unlike standard quantum field theory where we have an infinite number of degrees of freedom at each spacetime point.  This is supported by Bekenstein's argument, the black hole calculations in both string theory  and loop quantum gravity and is related to holographic ideas.  

It has been suggested that such a locally finite version of quantum field theory should be implemented by a many-Hilbert space theory (as opposed to the single Fock space for the entire universe in quantum field theory).  A QCH on a causal set is exactly such a locally finite quantum field theory.  This can be seen most clearly by formulating QCH as a locally finite analogue of algebraic quantum field theory.  
 Algebraic quantum field theory
 is a general approach to quantum field theory based on algebras of local observables, the relations among them, and their representations  (Haag 1992). 
 A QCH provides a similar discrete version as follows.

Let $\Gamma$ be a causal set.  This is a
partial order of events, the locally finite analogue of a Lorentzian spacetime.  Two events are causally related when $x\leq y$ and spacelike otherwise.  A parallel set $\xi$ is  the discrete analogue of a spacelike slice or part of a spacelike slice.  The causal relation $\leq$ is transitive.  

An algebraic quantum field theory associates a von\,Neumann algebra to each causally complete region of spacetime. This generalizes easily to a directed graph.  
The following definitions are exactly the same as for continuous spacetime. For any subset $X\subset \G$, define the \emph{causal complement} as
\[
X' := \{y\in\G \mid \forall x\in X : x\sim y\}
\]
the set of events which are spacelike to all of $X$. The \emph{causal completion} of $X$ is $X''$, and $X$ is \emph{causally complete} if $X=X''$. A causal complement is always causally complete (i.e., $X'''=X'$).

In the most restrictive axiomatic formulation of algebraic quantum field theory there is a von\,Neumann algebra $\A(X)$ for every causally complete region.  These all share a common Hilbert space. Whenever $X\subseteq Y$, $\A(X)\subseteq\A(Y)$. For any causally complete region $X$, $\A(X')$ is $\A(X)'$, the commutant of $\A(X)$. The algebra associated to the causal completion of $X\cup Y$ is generated by $\A(X)$ and $\A(Y)$\footnote{Some of the standard arguments about the properties of the local von\,Neumann algebras are valid for causal sets; some are not. The algebras should all be simple (i.e., von\,Neumann factors) because the theory would otherwise have local superselection sectors. For continuous spacetime it is believed that the local algebras should be type III${}_1$ hyperfinite factors; however, the reasoning involves the assumption that there exists a good ultraviolet scaling limit. This does not apply here; the small-scale structure of a causal set is discrete and not self similar at all.
}.

In our discrete version,  only a finite amount of structure should be entrusted to each event. In other words, each von\,Neumann algebra should be a finite-dimensional matrix algebra. In von\,Neumann algebra terms, these are finite type I factors.
Not surprisingly, simple matrix algebras are much easier to work with than type III von\,Neumann factors. Using the (unique) normalized trace, any state is given by a density matrix. Recall that the adjoint maps $\m^\dagger(x,y)$ in a quantum causal history are the induced maps on density matrices.

So, we see that the obvious notion of an algebraic quantum field theory on a causal set, with the physically reasonable assumption of finite algebras on events, gives the structure of a QCH. 
This means that the structure of a QCH encompasses a reasonable notion of a quantum field theory, and hence is capable of describing matter degrees of freedom.

 This framework may be a good one to investigate questions such as the transplanckian mode problem that arises in attempts at a locally finite quantum field theory in an expanding universe (for example in Foster and Jacobson, 2004).  
For the purposes of quantum gravity, this is a background {\em dependent} theory:  $\G$ is fixed, we only follow the dynamics of the $\A(x)$'s on the $\G$ which does not affect $\G$ itself.

%%%%%%%%%%%%%%%%%%%%%%%%%%%%%%%%%%%%%%%%%%%%%%

\section{Background independent theories of
quantum geometry}
\label{QG}
 
The traditional  path to a background independent candidate quantum theory of gravity is to consider a quantum superposition of geometries.  This is the case in Loop Quantum Gravity, quantum Regge calculus and causal sets and more recently spin foams and Causal Dynamical Triangulations.  

These are realizations of BI-II theories and can be illustrated by 
a QCH in a straightforward way:
$\Gamma$ will be a causal set, namely, a partial order of events that are causally related when $x\leq y$ and spacelike when $x\sim y$.  
To each event $x$, we shall associate an elementary  space of quantum geometrical degrees of freedom that are postulated to exist at Planck scale.  The theory provides a {\em sum}-over-all $\G$ amplitude to go from an initial to a final quantum geometry state.  
For example, this can be done as in {\em causal spin networks} (Markopoulou and Smolin, 1997; Markopoulou, 1997).  

Spin networks are graphs with directed edges labeled by representations of SU(2). 
  Reversing the direction of an edge means taking the conjugate representation.  A node in the graph represents the possible channels from the tensor product of the representations $\rho_{e_{\rm in}}$ on the incoming edges $e_{\rm in}$ to the tensor product of the representations on the outgoing ones,
  $\rho_{e_{\rm out}}$, i.e., it is the linear map
\beq
\iota:\bigotimes_{e_{\rm in}} \rho_{e_{\rm in}}\rightarrow\bigotimes_{e_{\rm out}} \rho_{e_{\rm out}}.  
\eeq
Such a map $\iota$ is called an {\em intertwiner}.
The intertwiners on a node form a finite-dimensional vector space. 
Hence, a subgraph in the spin network containing one node $x$ 
 corresponds to a Hilbert space $\Hi(x)$ of intertwiners.    
 Two 
spacelike events are two independent subgraphs, and the joint Hilbert 
space is $\Hi(x\cup y)=\Hi(x)\otimes \Hi(y)$ if they have no common edges, 
or $\Hi(x\cup y)=\sum_{\rho_{1},\ldots \rho_{n}}\Hi(x)\otimes \Hi(y)$, if $x$ and $y$  are joined in the spin network graph by 
$n$ edges carrying representations $ \rho_1,\ldots \rho_{n}$.    

Given an initial spin network, to be thought of as modeling a quantum
``spatial slice'', $\Gamma$ is built by repeated application of
local moves, local changes of the spin network graph.  Each move 
is a causal relation in the causal set.    The standard set of local generating moves for 4-valent spin networks is given by the following four operators:
\[
\begin{array}{rccc}
A_1=&\begin{array}{c}\includegraphics{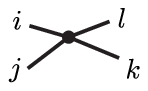}\end{array}&
\longrightarrow&
\begin{array}{c}\includegraphics{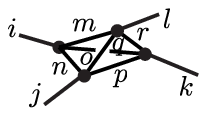}\end{array}
\\
A_2=&\begin{array}{c}\includegraphics{H42.eps}\end{array}&
\longrightarrow&
\begin{array}{c}\includegraphics{H41.eps}\end{array}
\\
%\end{array}
%\]
%\newpage
%\[
%\begin{array}{rccc}
%
A_3=&\begin{array}{c}\includegraphics{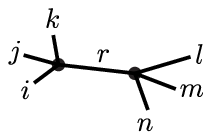}\end{array}&
\longrightarrow&
\begin{array}{c}\includegraphics{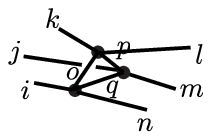}\end{array}\\
A_3=&\begin{array}{c}\includegraphics{H44.eps}\end{array}&
\longrightarrow&
\begin{array}{c}\includegraphics{H43.eps}\end{array}
\end{array}
\]
Note that the new subgraph has the same boundary as the original one and therefore corresponds to the same vector space of 
intertwiners\footnote{
Spin networks were originally
defined by Penrose as trivalent graphs with edges labelled by
representations of $SU(2)$.   Later, in Loop
Quantum Gravity, spin networks were shown to be the basis states for
the spatial geometry states.  The kinematical quantum area and volume operators,
in the spin network basis, have discrete spectra, and their
eigenvalues are functions of the labels on the spin network. }
\footnote{One uses 4-valent networks and moves for $SU(2)$ spin networks, instead of the simpler 3-valent ones we used in Fig.\ref{fig:moves} because 3-valent $SU(2)$ intertwiner spaces are one-dimensional and thus trivial.  

Also note that there is no preferred foliation in this model. 
The allowed moves change the network locally and any foliation
consistent with the causal set (i.e.\ that respects the order the
moves occured) is possible.  This is a discrete analogue of multifingered time evolution.  For more details, see
Markopoulou, 1997.
}.  A move $A_i$ is a unitary operator from a state 
$|S\rangle$ to a new one $|S'\rangle$ in $\Hi$. 

 A path integral quantum theory of gravity is then  obtained from the superposition of all possible $\G$'s, leading to an amplitude of the form 
\beq
A_{S_{\rm in}\rightarrow S_{\rm out}}=\sum_{\partial \Gamma = S_{\rm in}\rightarrow 
S_{\rm out}}\qquad
\prod_{{\mbox{\footnotesize moves}}\in\Gamma} A_i({\mbox{{move}}})
\label{eq:Zcsn}
\eeq 
to go from initial spin network $S_{\rm in}$ to final spin network $S_{\rm out}$.

%%%%%%%%%%%%%%%%%%%%%%%%%%%%%%%%%%%%%%%%%%%%%%

\subsection{Advantages and Challenges of quantum geometry theories.}\label{ACQG}

Particular realizations of quantum geometry theories, such as Loop Quantum Gravity, spin foams or CDT, amount to quantizations, canonical or path integral,  of General Relativity.  A specific quantization procedure will result in specific elementary state spaces and evolution operators.  The advantage of this is clear:  one follows the well-tested path to a new theory via the quantization of the classical one, a method that has been successful with all other theories that we have tried.  

Nonetheless, progress has been difficult, precisely because of the background independence of the classical theory, a feature that distinguishes it from all other theories that we have successfully quantized.  The equations of General Relativity are invariant under the diffeomorphism group of the manifold under investigation.  A canonical analysis reveals that this means that the system is completely constrained:  instead of generating time evolution, the Hamiltonian vanishes on solutions.  That means that in the description above, any intuition we may have of the $\G$ as describing changes of the network in time is incorrect, instead it represents a {\em projector} from the kinematical spin network states to the physical solutions.  This fact makes it especially hard to tackle questions of physical importance such as the emergence of the classical low energy limit, i.e., the recovery of the classical theory from the quantum gravity candidate.   

Without going in detail into specific issues that arise in each of the BI approaches to quantum gravity, one can get an idea of the problems that one encounters in the quest for the low energy limit of background independent theories, especially issues specific to BI systems by comparing our example to a condensed matter system.  
The graph $\G$  plays the role of the lattice, while the ${\Hc}_n$'s are the microscopic quantum degrees of freedom.  The low energy problem is analogous to describing the macroscopic behaviour emergent from a many-body system in condensed matter physics.  Building on that analogy, there has been work, for example, on the application of renormalization group methods to such BI systems (Markopoulou 2000; Oeckl 2002; Livine and Oriti,  2005; Manrique et al, 2005; Bombelli et al., 2005).  

There are, of course, technical obstacles such as the irregular nature of the lattices, the often complicated calculations involving the microscopic variables (usually group representations) and the lack of experimental controls,  readily available in standard condensed matter systems.  But there are also problems specific to BI systems:

\begin{itemize}

\item
{\em Dynamics}.
The low energy behavior of a physical  system depends on its dynamics.  

Causal Dynamical Triangulations (CDT)  is a clear demonstration of this basic fact of physical systems in quantum gravity.  Both CDT and Euclidean Dynamical Triangulations  (DT)  start with building blocks of the same dimensionality, four-simplices.  They differ in the dynamics.  In the continuum limit, CDT finds Hausdorff and heat dimensions near 3+1, while the Euclidean theory ends up either with effective dimension of 2 or infinite.  
Dynamics is notoriously difficult to implement in most background independent approaches, which makes it tempting to draw conclusions about the physical content of a theory before we have taken dynamics into account.  For example, spin foam models often relate the valence of the nodes in the spin foam 2-complex to the dimensionality of the system and much of the analysis of specific models involves analyzing the properties of a single building block  without considering the entire path-integral.  This is analogous to considering a spin system in condensed matter physics and inferring properties of its continuum limit by looking at the spins,  independently of the hamiltonian.  The Ising model in 2 dimensions and string networks (Wen 2005) have precisely the same building blocks and kinematics, square lattices of spins, but different dynamics.  The resulting effective theories could not be more different.   In the field of quantum gravity itself, the example of CDT vs DT shows us how little trust we should put in properties of the microscopic constituents surviving to the low-energy theory.  

We must conclude that any method we may use to analyze the low-energy properties of a theory needs to take the dynamics into account.

\item
{\em Observables}.  
Using the analogy between the graphs $\G$ of our theory and a condensed matter system, we may consider applying condensed matter methods to the graphs, such as a real space renormalization (coarse-graining the graph).  However, careful inspection of the real space renormalization method in ordinary systems shows that 
implicit in the method is the fact that, coarse-graining the lattice spacing coarse-grains the observables.  
In BI systems, the best we can do is relational observables and there is 
 no direct relationship between BI observables and the lattice or the history.  Hence,  the physical meaning of coarse-graining a graph is unclear. 
 
 In  theories of regularized geometries, such as CDT, there is a somewhat different issue.  The continuum limit observables that have been calculated so far are averaged ones, such as the Hausdorff or heat dimensions.  One still needs to find localized observables in order to compare the predictions of the theory to our world.  

\item
{\em (Lack of) symmetries}.
We should clarify that when we use the term {\em low-energy} it is only by analogy to ordinary physics and both {\em energy} and {\em low} are ill-defined.  The definition of energy needs a timelike Killing vector field, clearly not a feature of a BI theory.  A notion of scale is necessary to compare {\em low} to {\em high}.  Outside CDT, it is not clear how scale enters BI systems.

\end{itemize}

Note that all of the above issues are really different aspects of the question of dynamics in background independent theories.

%%%%%%%%%%%%%%%%%%%%%%%%%%%%%%%%%%%%%%%%%%%%%%

\section{Background Independent  pre-geometric systems}
\label{pregeo}

Is it possible to have a system that satisfies the definition of BI-I in section \ref{BI} but does not take the form of quantum geometry as in BI-II?  Even if this is possible, would such an object be of relevance in quantum gravity research?  The answer to both of these questions is not only yes but it constitutes an entire new direction in quantum gravity with a new set of exciting ideas.  

First, let us note that the example system of section \ref{ex} viewed as a quantum information processing system is BI-I in the obvious sense:  it describes a network of quantum systems and makes no reference to any spatiotemporal geometry.  More precisely, one can ask what 
 a quantum information processing system (a quantum computer) and our locally evolving networks have in common?  The answer is that they are the same mathematical structure, tensor categories of finite-dimensional vector spaces with arrows that are unitary or CP operators.   This is simply the mathematics of finite dimensional quantum systems.  What is interesting for us is that this mathematics contains no reference to any background spacetime that the quantum systems may live in and hence it is an example of  BI-I.

In the past two years, a number of BI-I systems have been put forward:  Dreyer's internal relativity (Dreyer, 2004 and this volume), Lloyd's computational universe (Lloyd, 2005), emergent particles from a QCH (Kribs and Markopoulou, 2005)	
and 
Quantum Graphity (Konopka, Markopoulou and Smolin, 2006).  All of these can be easily written as a QCH (with a single $\G$ and no geometric information on the state spaces, hence BI-I), so we shall continue the discussion in the more general terms of a pre-geometric QCH, just as it was defined in section \ref{QCH}.  

%%%%%%%%%%%%%%%%%%%%%%%%%%%%%%%%%%%%%%%%%%%%%
\subsection{The geometrogenesis picture}\label{sec:genesis}

Let us consider a simple scenario of what we may expect to happen in a BI theory with a good low energy limit.   It is a factor of about twenty orders of magnitude from the physics of the Planck scale  described by the microscopic theory to the standard subatomic physics.  By analogy with all other physical systems we know, it is reasonable to expect that physics at the two scales decouples to a good approximation.  We can expect at least one phase transition interpolating between the microscopic BI phase and the familiar one in which we see dynamical geometry.  We shall use the word {\em geometrogenesis} for this phase transition.

This picture implements the idea that spacetime geometry is a derivative concept and only applies in an approximate emergent level.  More specifically, this is consistent with the relational principle that spatial and temporal distances are to be defined internally, by observers inside the system.  This is the physical principle that led Einstein to special and general relativity.  The geometrogenesis picture implies that  the observers (subsystems), as well as any excitations that they may use to define such spatiotemporal measures, to be only applicable at the emergent geometric phase. 

The breakthrough realization (Dreyer, 2004 and this volume; Lloyd, 2005) is that the inferred geometry will necessarily be dynamical, since the dynamics of the underlying system will be reflected in the geometric description.  
This is most clearly stated by Dreyer who observes that since the same 
 excitations of the underlying system (characterizing the geometrogenesis phase transition) and their interactions will be used to define {\em both} the geometry and the energy-momentum tensor $T_{\mu\nu}$.   This leads to the following Conjecture on the role of General Relativity:
 
 \begin{quotation}
If the assignment of geometry and $T_{\mu\nu}$ from the same excitations and interactions is done consistently, the geometry and $T_{\mu\nu}$ will not be independent but will satisfy Einstein's equations as identities.
\end{quotation}

What is being questioned here is the separation of physical degrees of freedom into matter and gravitational ones.  In theories with a fixed background, such as quantum field theory, the separation is unproblematic, since the gravitational degrees of freedom are not really free and do not interact with the matter.  In the classical background independent theory, general relativity, we are left with an intricate non-linear relation between the two sets: the Einstein equations.  As  the practitioners of canonical quantum gravity know well, cleanly extracting dynamical gravitational degrees of freedom from the matter is fraught with difficulties.  If such a clean separation could be achieved, canonical quantum gravity would have succeeded at least two decades ago.  

The new direction unifies matter and gravity in the pre-geometric phase and provides a path towards {\em explaining} gravity rather than just quantizing it.  

%%%%%%%%%%%%%%%%%%%%%%%%%%%%%%%%%%%%%%%%%%%%%
\subsection{Advantages and Challenges of pre-geometric theories.}
\label{ACpregeo}

Such a radical move raises, of course, numerous new questions.  Due to the short time that this direction has been pursued, the advantages and the challenges here are not as well-studied as in the case of quantum geometry, we shall, however, list some here.  

The main advantage in practical terms is that this approach allows for ordinary quantum dynamics in the pre-spacetime theory, instead of a quantum constraint, potentially providing a way out of the issues listed in section \ref{ACQG}.  If successful, it promises a deeper understanding of the origin of gravity, usually beyond the scope of quantum geometry theories. 

The obvious challenges are: 
\begin{itemize}
\item
{\em Time.}
  Does the ordinary dynamics of the pre-geometric phase amount to a background time?  Keep in mind that there are strict observational limits on certain kinds of background time (Jacobson, Liberati and Mattingly, 2006).  
 Recent work indicates that the answer is not clear.  There are several possible mechanisms that may wipe out any signature of the pre-geometric time when we go through the phase transition  (Lloyd, 2005; Konopka and Markopoulou, 2006; Dreyer, this volume).  
 \item
{\em Geometry.}  How can we get geometry out if we do not put it in?  Presumably, most pre-spacetime systems that satisfy the QCH definition will not have a meaningful geometric phase.  Will we need a delicate fine-tuning mechanism to have a geometric phase or is there a generic reason for its existence?  

A variety of ways that geometry can arise have been proposed:  dispersion relations at the Fermi point (Volovik, also see Dreyer, this volume), symmetries of the emergent excitations (Kribs and Markopoulou, 2005), free excitations (Dreyer), restrictions on the properties of the graph $\G$ (Lloyd, 2005) or emergent symmetries of the ground state (Konopka, Markopoulou and Smolin, 2006).  It is promising that most of these point towards generic mechanisms for the presence of a regular geometry.  
\end{itemize}

%%%%%%%%%%%%%%%%%%%%%%%%%%%%%%%%%%%%%%%%%%%%%
\subsection{Conserved Quantities in a BI system}
\label{NS}

Admittedly, we only have guesses as to the microscopic theory and very limited access to experiment.  Additionally, phase transitions are not very well understood even in ordinary lab systems, let alone phase transitions of background independent systems. In spite of these issues, we find that the geometrogenesis picture suggests a first step towards the low energy physics that we can take.  

A typical feature of a phase transition is that the degrees of freedom that characterize each of the two phases are distinct (e.g.\  spins vs spin waves in a spin chain or atoms vs phonons in solid state systems), with the emergent degrees of freedom being collective excitations of the microscopic ones.  In our example, the vector spaces on graphs contain the microscopic degrees of freedom and operators in $\Aevol$ is the microscopic dynamics.  Is there a way to look for collective excitations of these that are long-range and coherent so that they play a role in the low-energy phase? 

We find that this is possible, at least in the idealized case of conserved (rather than long-range) quantities in a background independent system such as our example.   The method we shall use, {\em noiseless subsystems}, is borrowed from quantum information theory, thanks to the straightforward mapping between locally finite BI theories and quantum information processing systems which we described above. 
We are then suggesting a new path to the effective theory of a background independent system.  The basic strategy is to begin by identifying effective coherent degrees of freedom and use these and their interactions to characterize the effective theory.  If they behave as if they are in a spacetime, we have a spacetime.

In Kribs and Markopoulou, 2005, we found that the field of quantum information theory has a notion of coherent excitation which, unlike the more common ones in quantum field theory and condensed matter physics,  makes no reference to a background geometry and can be used on a BI system.  
This is the notion of a {\em noiseless subsystem} (NS) in quantum error correction, a subsystem protected from the noise, usually thanks to symmetries of the noise\footnote{
Zanardi and Rasetti, 1997;  Knill, Laflamme and Viola, 2000.}
.  Our observation is that passive error correction is analogous to problems concerned with the emergence and
stability of  persistent quantum states in condensed matter physics.     
In a quantum gravity context, the role of noise is simply the fundamental evolution and the existence of a noiseless subsystem 
means a coherent excitation protected from the microscopic
Planckian  evolution, and thus relevant for the effective theory. 

\begin{definition}{\em Noiseless Subsystems.}
Let $\Phi$ be a quantum channel on $\Hc$ and suppose that $\Hc$ decomposes as 
$\Hc =(\Hc^A\otimes\Hc^B)\oplus\K$, where $A$ and $B$ are subsystems and
$\K =(\Hc^A\otimes\Hc^B)^\perp$. We say that $B$ is {\em noiseless}
for $\Phi$ if
\begin{eqnarray}\label{ns}
\forall\sigma^A\ \forall\sigma^B,\ \exists \tau^A\ :\
\Phi(\sigma^A\otimes\sigma^B) = \tau^A\otimes \sigma^B.
\end{eqnarray}
\label{eq:NS}
\end{definition}
Here we have written $\sigma^A$ (resp.\ $\sigma^B$) for operators
on $\Hc^A$ (resp.\ $\Hc^B$), and we regard $\sigma = \sigma^A\otimes
\sigma^B$ as an operator that acts on $\Hc$ by defining it to be
zero on $\K$.  

In general, given $\Hi$ and $\Phi$, it is a non-trivial problem to find a decomposition that exhibits a NS.  Much of the relevant literature in quantum information theory is concerned with algorithmic searches for a NS given $\Hi$ and $\Phi$.  
However,  if we apply this method to the example theory of \ref{ex}, it is straightforward to see that it has a large conserved sector\footnote{
The noiseless subsystem method (also called decoherence-free
subspaces and subsystems) is the fundamental passive technique for
error correction in quantum computing.  In this setting, the operators $\Phi$ are
called the {\em error} or {\em noise} operators associated with
$\Phi$. It is precisely the effects of such operators that must be
mitigated for in the context of quantum error correction.
The basic idea in this setting is to (when
possible) encode initial states in sectors that will remain immune
to the deleterious effects of the errors $\Phi$
associated with a given channel.

The term ``noiseless'' may be confusing in the present context:  it is not necessary that there is a noise in the usual sense of a given split into system and environment.  As is clear from the definition above, simple evolution of a dynamical system is all that is needed, the noiseless subsystem is what evolves coherently under that evolution.}:

\begin{quote}

{\em Noiseless subsystems in our example theory.}

Are there any non-trivial noiseless subsystems in $\Hi$?   There are, and they are revealed when we rewrite $\Hi_S$ in eq.\ (\ref{eq:H}) as
\beq
\Hi_S=\Hi_S^{n'}\otimes\Hi_S^B,
\label{eq:Hnew}
\eeq
where $\Hi_S^{n'}:=\bigotimes_{n'\in S}\Hi^{n'}$ contains all {\em unbraided} single node subgraphs in $S$ (the prime on $n$ serves to denote unbraided) and $\Hi_S^b:=\bigotimes_{b\in S}\Hi_b$ are state spaces associated to braidings of the edges connecting the nodes.  For the present purposes, we do not need  to be explicit about the different kinds of braids that appear in $\Hi_S^b$.  

The difference between the decomposition (\ref{eq:H}) and the new one (\ref{eq:Hnew}) is best illustrated with an example (details can be found in Bilson-Thompson, Markopoulou and Smolin, 2006).  Given the state
\beq
    \begin{array}{c}\mbox{\includegraphics{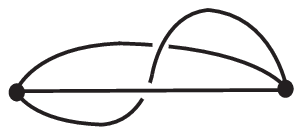}}\end{array}
\eeq
eq.\ (\ref{eq:H}) decomposes it as 
\beq
    \begin{array}{c}\mbox{\includegraphics{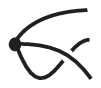}}\end{array}\otimes
     \begin{array}{c}\mbox{\includegraphics{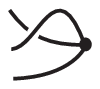}}\end{array}
\eeq
while (\ref{eq:Hnew}) decomposes it to 
\beq
       \begin{array}{c}\mbox{\includegraphics{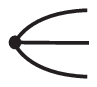}}\end{array}\otimes
    \begin{array}{c}\mbox{\includegraphics{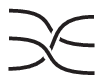}}\end{array}\otimes
     \begin{array}{c}\mbox{\includegraphics{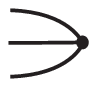}}\end{array}.
\eeq

With the new decomposition, one can check that operators in $\Aevol$ can only affect the $\Hi_S^{n'}$ and that $\Hi_S^b$ is {\em noiseless under} $\Aevol$.  This can be checked explicitly by showing that the actions of braiding of the edges of the graph and the evolution moves commute.  

We have shown that braiding of graph edges are unaffected by the usual evolution moves.  Any physical information contained in the braids will propagate coherently under $\Aevol$.  These are effective coherent degrees of freedom\footnote{The physical interpretation of the braids is beyond the scope of this paper.  See Bilson-Thompson, Markopoulou and Smolin, 2006,  for an interpretation of the braids as quantum numbers of the standard model.
}.  

Note that this example may appear simple but the fact that the widely used system of locally evolving graphs exhibits broken ergodicity  ($\Hc$ splits into sectors, characterized by their braiding content, and $\Aevol$ cannot take us between sectors) went unoticed prior to the introduction of the NS method.  
\end{quote}

Before closing, we would like to point out some of the subtleties of background independence that, not surprisingly, arise here.  Our original motivation to search for conserved quantities was that they can be thought of as a special case of emergent long-range propagating degrees of freedom, where the lifetime of the propagating ones is infinite (and so tell us something about the geometric phase of the theory).  Noiseless subsystems can only deal with this case because it only looks at the symmetries of the microscopic dynamics.  Presumably, what we need is to weaken the notion of a noiseless subsystem to ``approximately conserved'' so that it becomes long-range rather than infinite.  Long-range, however, is a comparative property and to express it we need a way to introduce scale into our system.  It is unclear at this point whether it is possible to introduce a scale in a pre-geometric theory without encountering the problems listed in section \ref{ACQG}.

%%%%%%%%%%%%%%%%%%%%%%%%%%%%%%%%%%%%%%%%%%%%%
\section{Summary and Conclusions}
\label{conc}

In this article we started with the traditional background independent approaches to quantum gravity which are based on quantum geometric/gravitational degrees of freedom.   We saw that, except for the case of causal dynamical triangulations, these encounter significant difficulties in their main aim, i.e., deriving general relativity as their low energy limit.  We then suggested that general relativity should be viewed as a strictly effective theory coming from a fundamental theory with no geometric degrees of freedom (and hence background independent in the most direct sense).  

The basic idea is that an effective theory is characterized by effective coherent degrees of freedom and their interactions. Having formulated the pre-geometric BI theory as a quantum information theoretic processor, we were able to use the method of noiseless subsystems to extract such coherent (protected) excitations. 

The geometrogenesis picture leads one to reconsider  the role of microscopic quantum geometric degrees of freedom traditionally present in background independent theories.  It appears unnatural to encounter copies of the geometry characteristic of the macroscopic phase already present in the microscopic phase, as is the case, for example, when using quantum tetrahedra in a spin foam.  Instead, one can start with a pre-geometric theory and look for the effective coherent degrees of freedom along the lines described.  Spacetime is to be inferred  by them internally, namely, using only operations that are accessible to parts of the system.  

This  is very promising for three reasons:  1) The emphasis on the effective coherent degrees of freedom addresses directly and in fact uses the dynamics.  The dynamics is physically essential but almost impossible to deal with in other approaches.  2) A truly effective spacetime has novel phenomenological implications not tied to the Planck scale which can be tested and rejected if wrong.  3) A pre-spacetime background independent quantum theory of gravity takes us away from the concept of a quantum superposition of spacetimes which can be easily written down formally but has been impossible to make sense of physically in any approach other than Causal Dynamical Triangulations.  

Some of the more exciting possibilities we speculated on included solving the problem of time and {\em deriving} the Einstein equations.  Clearly this direction is in its beginning, but the basic message is that taking the idea that general relativity is an effective theory seriously involves rethinking physics without spacetime.  
This opens up a whole new set of possibilities and opportunities.

\end{document}